\documentclass{PoS}
\usepackage{amsmath,slashed,graphicx}
\usepackage[leftcaption]{sidecap}

\renewcommand{\dag}{\dagger}
\newcommand{\g}{\gamma}
\renewcommand{\a}{\alpha}
\renewcommand{\b}{\beta}

\title{The Neutrino Option}

\ShortTitle{The Neutrino Option}

\author{\speaker{Ilaria Brivio}\thanks{Supported by the Villum Fonden and the Danish National Research Foundation (DNRF91).}\\
        Niels Bohr International Academy and Discovery Center, Niels Bohr Institute,
University of Copenhagen, Blegdamsvej 17, DK-2100 Copenhagen, Denmark\\
        E-mail: \email{ilaria.brivio@nbi.ku.dk}}

\abstract{We discuss the possibility that the Higgs potential and electroweak scale are generated radiatively in a type-I seesaw scenario. 
A Higgs potential consistent with experimental constraints can be obtained in this hypothesis for a Majorana mass scale $m_N\sim 10-500$~PeV and with neutrino Yukawa couplings of order $|\omega|\sim 10^{-4.5}-10^{-6}$. Remarkably, neutrino masses and mixings can be simultaneously accommodated within this parameter space.
This framework, that ties together Higgs phenomenology, precision top quark mass measurements and neutrino physics, represents an alternative approach to the hierarchy problem, in which the Higgs mass is not stabilized around the TeV scale, but rather determined by radiative corrections at higher energies. Traditional hurdles in overcoming the hierarchy problem are then traded for the new challenge of generating PeV Majorana masses while suppressing the tree-level scalar potential in the UV.

}

\FullConference{The European Physical Society Conference on High Energy Physics\\
		5-12 July, 2017\\
		Venice}

\begin{document}

\section{Introduction}

One of the major lacunas of the Standard Model (SM) is the lack of a dynamical origin for the Higgs potential, that renders the vacuum structure of the theory extremely sensitive to  quantum corrections from heavy states coupling to the field $H$. 
These states generally give finite radiative contributions to $(H^\dag H)$, that are proportional to the square of their mass and remain present when the heavy particles are integrated out of the spectrum. This directly follows from dimensional analysis and considering the matching of a generic UV Lagrangian onto the SM in an Effective Field Theory (EFT) approach.
Using the EFT language, here we refer to those quantities as threshold matching contributions. The hierarchy problem can then be expressed in terms of thresholds $\Delta m_h^2\gg (125$~GeV$)^{2}$ being typically induced in the presence of heavy new physics sectors.

The most popular solutions to this issue aim at suppressing these contributions: they typically use symmetries either to enforce cancellations among different threshold contributions (for instance supersymmetry) or to make the thresholds naturally small and close to the electroweak (EW) scale (e.g. the shift symmetry typical of composite Higgs models). A common feature of these scenarios is that, by construction, the potential is stabilized at the TeV scale.
This point appears to be problematic for two main reasons: one is that it implies that new particles are expected close to the TeV energy range, a condition that has not been yet validated by experimental evidence. The other reason is that the measured Higgs mass and vacuum expectation value point to a relatively small quartic coupling $\lambda\sim 0.13$, a value that is quite difficult to obtain at the TeV scale without introducing additional fine tunings~\cite{Bellazzini:2014yua}. The latter are inevitable if the potential is tree-level generated, which typically yields a larger $\lambda$. On the other hand, tensions in the parameter space also emerge assuming that the potential fully originates at one-loop in a weakly coupled theory.
For instance, in composite Higgs models this assumption would be natural for a compositeness scale $f\sim v$, which is in contrast with the measurements of the Higgs couplings and with EW precision tests, that instead point to $f>v$~\cite{Bellazzini:2014yua}. For these reasons, it is worth considering alternative approaches.

\begin{SCfigure}[.55][b]
 \includegraphics[width=10cm]{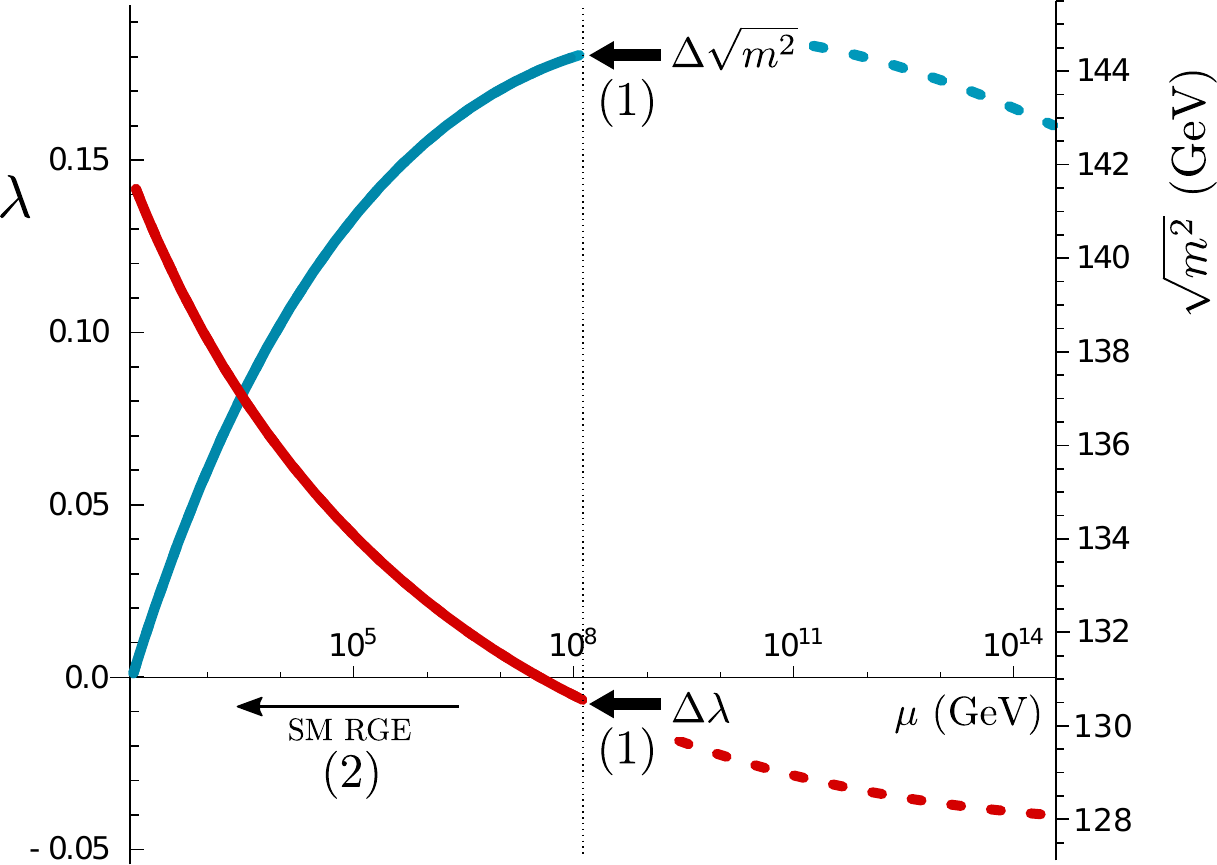}
\caption{Schematic illustration of the approach presented: the parameters of the Higgs potential are first (1) fixed by threshold corrections $\Delta\lambda,\, \Delta\sqrt{m^2}$ at a scale much higher than the TeV and then (2) run down to the EW scale according to the SM RGE. The red line and left axis (blue line and right axis) show $\lambda(\mu)$ ($\sqrt{m^2(\mu)}$) according to the SM 1-loop running~\cite{Buttazzo:2013uya}.

\vspace*{8mm}
}\label{fig:main_idea}
\end{SCfigure}

An interesting possibility is to employ threshold corrections, rather than suppressing them, to generate the scalar potential at a scale $M$ (much) larger than the TeV~\cite{Brivio:2017dfq}. 
The idea, illustrated schematically in Fig.~\ref{fig:main_idea}, is that the potential parameters can be determined by the thresholds induced  at energies $\mu\simeq M$; their values would then run down to the EW scale according the SM RGE, that are known~\cite{Buttazzo:2013uya}.
Generating the scalar potential at higher energies is convenient both because the constraints from direct searches are easily evaded and because the relations among the parameters are altered due to running effects, which can make the construction of the potential numerically easier. A particularly interesting region is, for instance, where $\lambda\simeq 0$: as we show below, the potential can be produced at this scale by a new physics sector giving a suppressed $(H^\dag H)^2$.
A major assumption in this novel setup is that, in the UV, the bare tree-level potential should be subdominant  compared to the thresholds. Realizing this condition in a concrete model represents the counterpart of the usual challenges due to the hierarchy problem for this new perspective.

In this talk we show how this idea can be successfully applied in the (arguably) simplest extension of the SM, the type-I seesaw model, obtaining realistic values for the Higgs potential and for the absolute neutrino mass scale $m_\nu$ simultaneously \cite{Brivio:2017dfq}.

\section{Threshold corrections from the seesaw model}
We consider a standard type-I seesaw scenario~\cite{Minkowski:1977sc,GellMann:1980vs,Mohapatra:1979ia,Yanagida:1980xy}, extending the SM  with 3 singlet right-handed fields $N_{R,p}$, $\,p=\{1,2,3\}$. In the basis in which the mass matrix $m_p$ is real and diagonal, the relevant Lagrangian is (we use the notation of~\cite{Broncano:2002rw,Elgaard-Clausen:2017xkq})
\begin{equation}
2 \,\mathcal{L}_{N_p}=  \overline{N_p} (i\slashed{\partial} - m_{p})N_p - \overline{\ell_{L}^\beta} \tilde{H} \omega^{p,\dagger}_\beta  N_p
-  \overline{\ell_{L}^{c \beta}} \tilde{H}^* \, \omega^{p,T}_\beta N_p - \overline{N_p} \, \omega^{p,*}_\beta \tilde{H}^T \ell_{L}^{c \beta}  - \overline{N_p}\, \omega^p_\beta \tilde{H}^\dagger \ell_{L}^\beta,
 \label{LN}
\end{equation}
where $\ell$ is the SM lepton doublet, with flavor index $\beta$, and the superscript $c$ denotes charge conjugation defined as $\psi^c = -i\g_2\g_0 \bar\psi^T$.
The fields $N_p$ satisfy the Majorana condition $N_p^c=N_p$
and they are related to the chirality eigenstates by {$N_p = e^{i\theta_p/2} N_{R,p}+ e^{-i\theta_p/2}(N_{R,p})^c$}, being $\theta_p$ the Majorana phases~\cite{Broncano:2002rw}.
The couplings $\omega_\beta^p=\{x_\beta,y_\beta,z_\beta\}$ are complex vectors in flavor space. Finally, here and in the following, repeated indices are summed over, unless otherwise stated.

Integrating out the heavy eigenstates $N_p$ and matching onto the dimension 5 Weinberg operator gives the well-known Majorana mass matrix for the 3 light neutrinos  $(m_\nu)_{\a\b} = v^2(\omega_\a^p)^T \omega_\beta^p /2m_p$.

\begin{figure}[t]\centering
 \includegraphics[width=.7\textwidth]{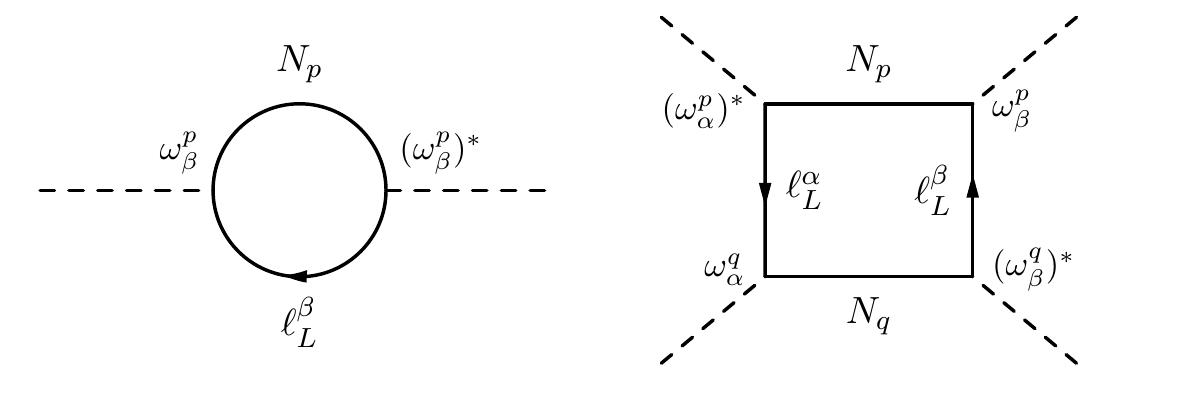}
 \caption{One-loop diagrams leading to threshold corrections to the Higgs potential in the seesaw model.}\label{fig:diagrams}
\end{figure}
Threshold corrections to the Higgs potential are induced in the seesaw model due to the diagrams in Fig.~\ref{fig:diagrams}. The one-loop finite terms are found to be\footnote{The loops are calculated using dimensional regularization and $\overline{\rm MS}$ renormalization. The threshold corrections are then identified as the zero-th order term in the expansion in $v/M$ (see e.g. Ref.~\cite{Brivio:2017vri} for a detailed discussion).}~\cite{Brivio:2017dfq}
\begin{equation}\label{DeltaV}
\begin{aligned}
\Delta V(H^\dagger \, H) =& -\frac{m_p^2 |\omega_p|^2}{16 \, \pi^2}\,\Bigg[ 1 + \log \frac{\mu^2}{m_p^2}\Bigg] \, H^\dagger H\\[-2mm]
&
- \frac{5 \, (\omega_q \cdot \omega^{p,\star})(\omega_p \cdot \omega^{q,\star})}{64 \,\pi^2} \, \Bigg[1 - \frac{m_p m_q \, \log \frac{m_p^2}{m_q^2} + m_q^2 \log \frac{\mu^2}{m_q^2}-m_p^2
\log \frac{\mu^2}{m_p^2}}{(m_p^2-m_q^2)} \,\Bigg] \, (H^\dagger H)^2.
\end{aligned}
\end{equation} 
in agreement with~\cite{Casas:1999cd,Bambhaniya:2016rbb}.
\begin{figure}[t]
 \includegraphics[width=.49\textwidth]{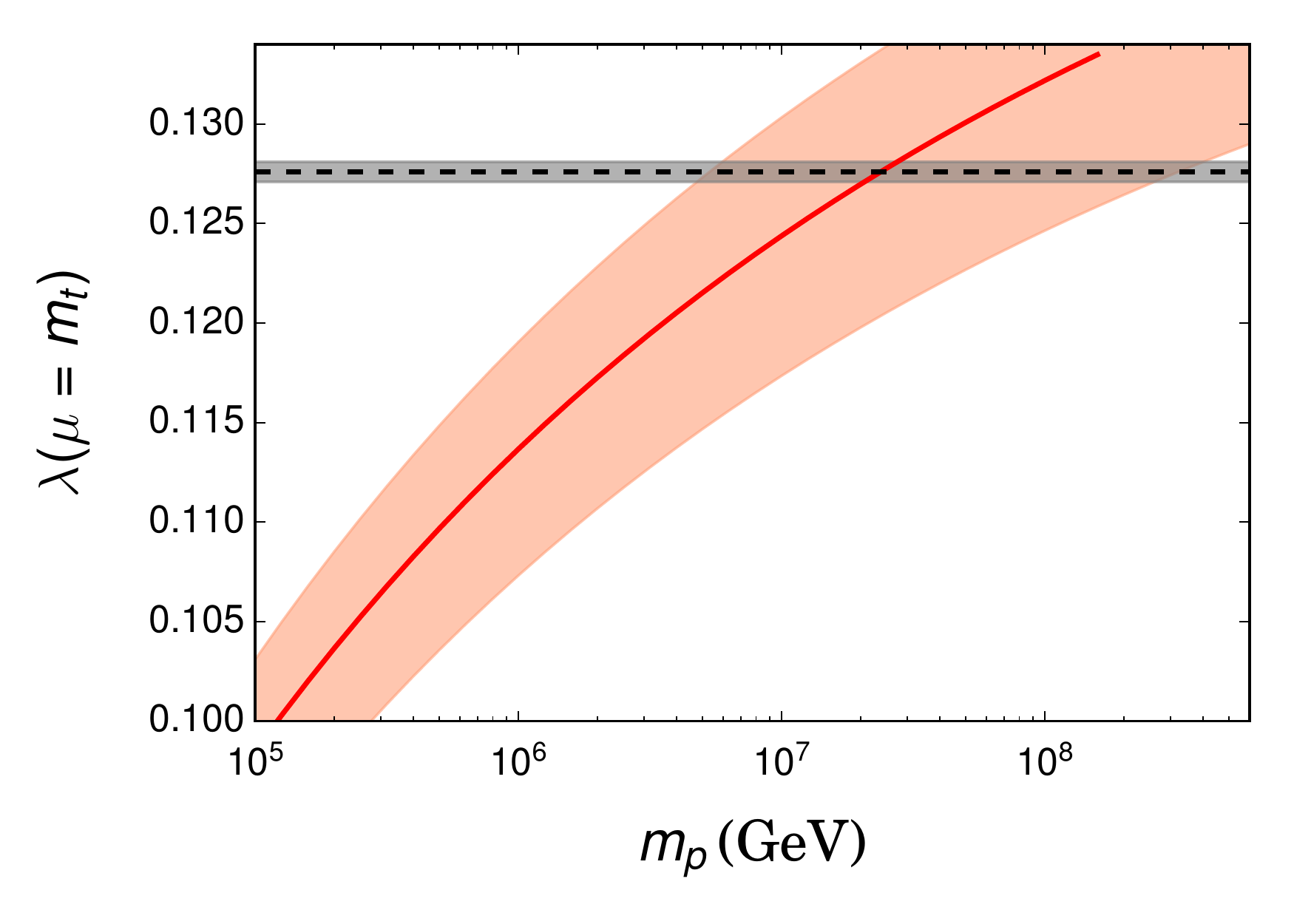}
 \includegraphics[width=.49\textwidth]{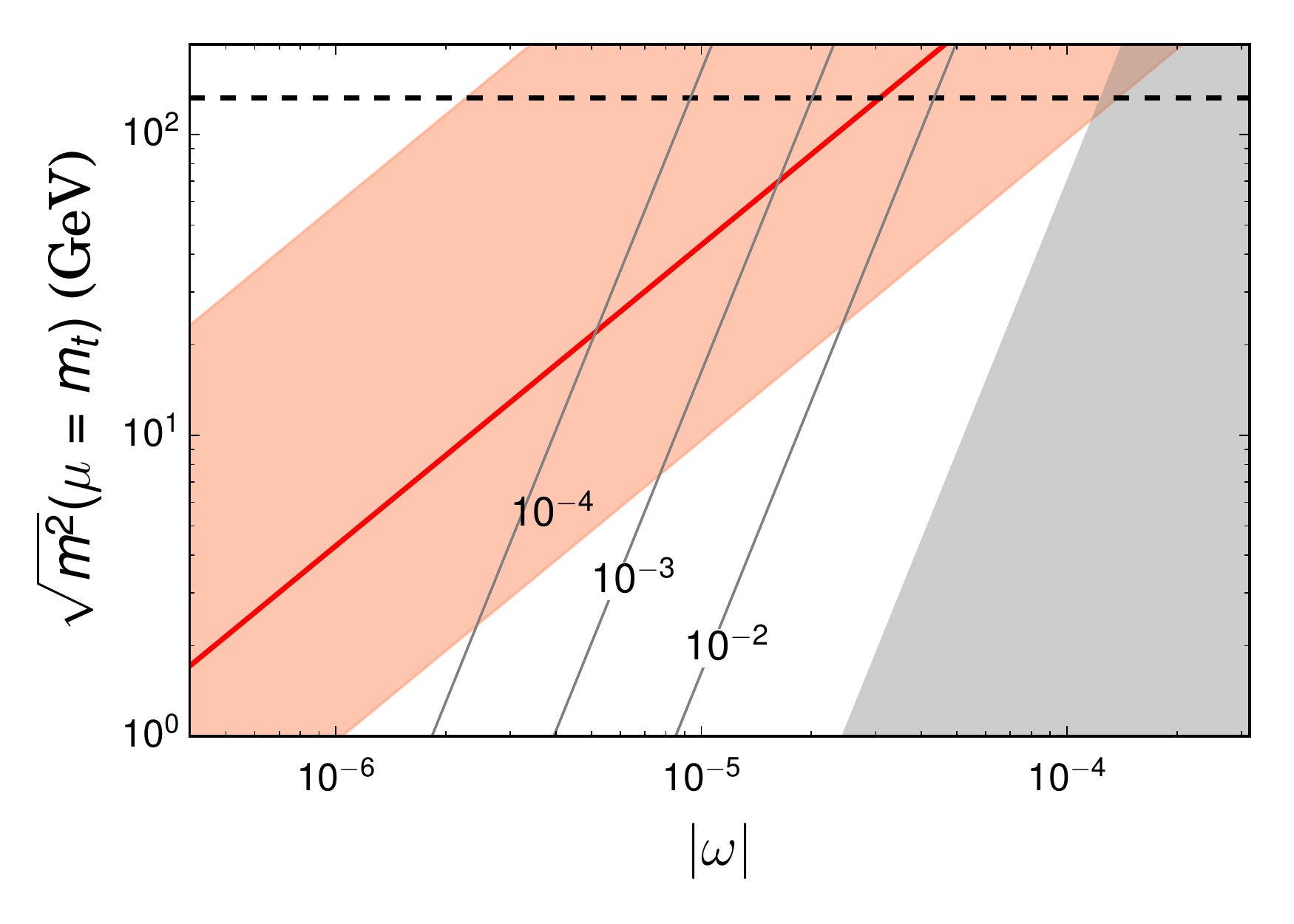}
 \caption{Values of the parameters $\lambda$ (left) and $\sqrt{m^2}$ (right) extrapolated at the scale $\mu=\hat{m}_t$ as a function of the heavy neutrino mass scale $m_p$ and of $|\omega|$ respectively. The dashed lines and surrounding bands indicate the values consistent with the measured Higgs mass within $\pm 1\sigma$~\cite{Aad:2015zhl}. The red line in the left panel assumes $\hat{m}_t=173.2$~GeV and the surrounding band corresponds to varying $\hat{m}_t$ between 171 and 175~GeV (a $2 \sigma$ error variation \cite{Olive:2016xmw}).
 In the right panel, the solid red line assumes $m_p = 10^{1.3}$~PeV, and the grey region is disfavoured due to $\Lambda$ CDM cosmology limit $\sum m_\nu \leq 0.23$~eV. The three solid lines indicate, for reference, the sum of neutrino masses predicted, in ${\rm eV}$.}\label{fig:running}
\end{figure}
Assuming that the Majorana sector has a nearly degenerate spectrum ($(m_q^2-m_p^2)/m_{p,q}^2 < 1$) all the heavy masses can be approximated to a universal scale $m_p$. In this limit, both expressions in square brackets in Eq.~\eqref{DeltaV} reduce to 1 when evaluated at $\mu=m_p$.
Eq.~\eqref{DeltaV} can be further simplified, without loss of generality, reducing the entries of $\omega_\beta^p$ to a common coupling scale $|\omega|$. 
The threshold contributions are then:
\begin{equation}\label{thresholds}
\Delta \, m^2 = m_p^2 \frac{|\omega_p|^2}{8 \, \pi^2}, \qquad \Delta \, \lambda =-5 \frac{|\omega|^4}{64 \,\pi^2}\,,
\end{equation} 
having parameterized the classical Higgs potential as
$ V_c(H^\dagger H) = - (m^2 / 2) \,(H^\dagger \, H) + \lambda \, (H^\dagger \, H)^2
$.
We assume that these are the dominant contributions to the scalar potential when the $N_p$ are integrated out.
The scenario postulated then requires (i) an approximate scale invariance, broken only\footnote{Breaking effects of this invariance from SM quantum corrections (in particular QCD contributions) and from the renormalization of the Coleman-Weinberg potential are expected to be subdominant as long as $m_p^2 |\omega|^2 \gg \Lambda_{\rm QCD}^2$~\cite{Brivio:2017dfq}.
} by the Majorana masses $m_p$, and (ii) the bare tree-level Higgs potential to be negligible compared to the thresholds \eqref{thresholds} at $\mu=m_p$. Constructing a model that fulfills these conditions goes beyond the scope of this work and represents a challenge alternative to the usual hierarchy problem. It is interesting to observe, however, that this scenario ties the breaking of scale invariance to the violation of lepton number, which serves as a protective symmetry for the scalar potential.

\section{Running down to the scale {\boldmath $\mu = \hat m_t$}}

\begin{SCfigure}[.87][t]\centering
\includegraphics[width=.5\textwidth]{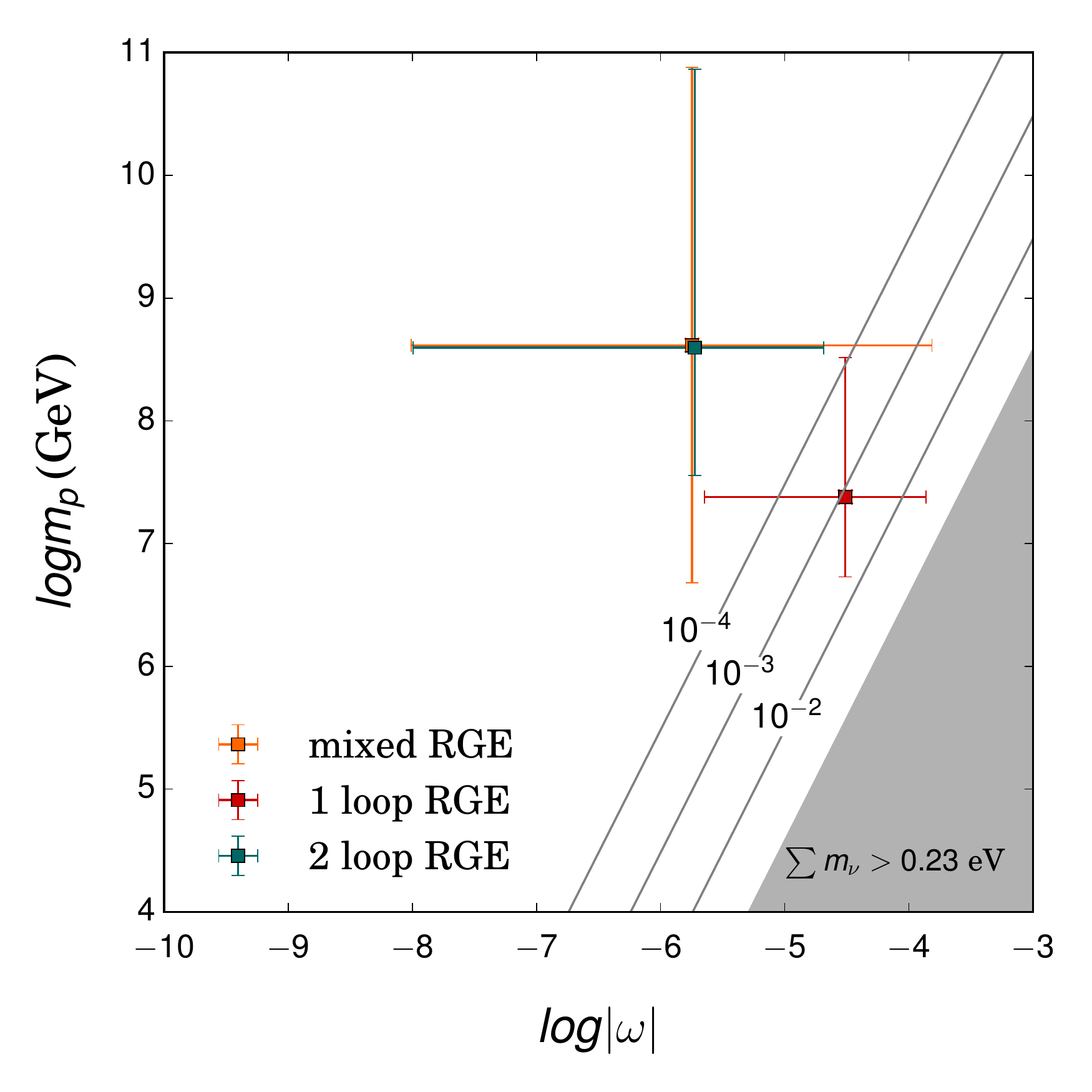}
 \caption{Numerical sensitivity of the results, with all cases showing one loop matching to $m^2$, $\lambda$ and including one loop corrections for the remaining SM parameters with one, two loop, or mixed RGEs for all parameters. The mixed case shows one loop running for $m^2$, $\lambda$ and two loop running for all remaining SM parameters. 
 The best fit points are indicated with a box in each case with error bars showing the experimental uncertainty in the top quark mass, taken to be its $2 \, \sigma$ uncertainty \cite{Olive:2016xmw}.
 \vspace*{2cm}
 }\label{fig:errors}
\end{SCfigure}

Below the scale $\mu=m_p$ the $N_p$ states are integrated out, and the parameters run according to the SM RGE. Here we consider the coupled system of 1-loop RGE for $m^2, \lambda$, the SM gauge couplings $g_1,\,g_2,\,g_3$ and the top Yukawa coupling $Y_t=\sqrt2 m_t/v$. Contributions from other Yukawas are neglected and the expressions of the $\beta$-functions are taken
from Ref.~\cite{Buttazzo:2013uya}. 
The differential system is solved fixing the boundary conditions 
$m^2(m_p) = \Delta m^2,\, \lambda(m_p) = \Delta \lambda,\,
 g_1(m_t)= 0.3668,\,  g_2(m_t) = 0.6390,\,
 g_3(m_t)=1.1671,\, Y_t(m_t) = 0.9460,$
where the value of $g_3$ includes RG running at 4-loops in QCD and 2-loops in the EW interactions~\cite{Buttazzo:2013uya} while those of the other SM parameters are computed at one-loop in the SM using the input parameters $G_F = 1.1663787\cdot 10^{-5}$~GeV$^{-2}$, {$\hat{\alpha}_s = 0.1185$},
$\hat m_Z = 91.1875$~GeV, $\hat m_W = 80.387$~GeV, $\hat m_h = 125.09$~GeV and $\hat m_t = 173.2$~GeV.

The values of $m^2(m_t)$ and $\lambda(m_t)$ are extrapolated from the RGE solutions as a function of $m_p$ and $|\omega|$. 
In particular, $\lambda(m_t)$ is found to depend significantly only on $m_p$ (see Fig.~\ref{fig:running}, left) and the correct value is obtained for $m_p\simeq 10^{1.3}$~PeV. This result shows a strong sensitivity to the top mass: varying $m_t$ within the $2\sigma$-allowed range 171--175 GeV (red band) can shift the optimal $m_p$ by about an order of magnitude.
Fig.~\ref{fig:running}, right,  shows the dependence of $\sqrt{m^2(m_t)}$ on $|\omega|$. 
The red solid line marks the curve with constant $m_p=10^{1.3}$~PeV, while the red band indicates the region spanned varying $m_p$ within the values that give the correct $\lambda(m_t)$ for $171\leq m_t\leq 175$~GeV. 
The results obtained also show a significant numerical dependence on the loop order used for the RGE, as illustrated in Fig.~\ref{fig:errors}.

For $m_t=173.2$~GeV and with 1-loop RGE, the correct EW scalar potential is generated with $m_p\simeq 10^{1.3}$~PeV and $|\omega|\simeq 10^{-4.5}$, pointing to neutrino masses such that $\sum m_\nu = 3|\omega|^2 v^2/2m_p\simeq 3\cdot 10^{-3}$~eV, which is within the window allowed by the cosmological constraint  $\sum m_\nu \leq 0.23$~eV~\cite{Ade:2015xua}.
Neutrino mass splittings and mixings are not reproduced in this simplified study, but they can in principle be accommodated, given that restoring the flavor structure of $m_p$ and $\omega_\beta^p$ introduces a number of free parameters that is larger than that of the experimental constraints. 
The scenario presented here is generally not compatible with thermal leptogenesis, because the expected size of the Yukawa couplings $|\omega|\sim 10^{-4.5}$ is too small to produce a sufficient amount of matter-antimatter asymmetry~\cite{Davoudiasl:2014pya}.
Finally, because the $N_p$ states are nearly decoupled, no other signature, beyond non-zero neutrino masses, is predicted to show up at the LHC or other near future experiments.

\providecommand{\href}[2]{#2}\begingroup\raggedright\endgroup


\begin{thebibliography}{10}

\bibitem{Bellazzini:2014yua}
B.~Bellazzini, C.~Cs\'aki and J.~Serra, \emph{{Composite Higgses}},
  \href{http://dx.doi.org/10.1140/epjc/s10052-014-2766-x}{\emph{Eur. Phys. J.}
  {\bf C74} (2014) 2766}, [\href{https://arxiv.org/abs/1401.2457}{{\tt
  1401.2457}}].

\bibitem{Buttazzo:2013uya}
D.~Buttazzo, G.~Degrassi, P.~P. Giardino, G.~F. Giudice, F.~Sala, A.~Salvio
  et~al., \emph{{Investigating the near-criticality of the Higgs boson}},
  \href{http://dx.doi.org/10.1007/JHEP12(2013)089}{\emph{JHEP} {\bf 12} (2013)
  089}, [\href{https://arxiv.org/abs/1307.3536}{{\tt 1307.3536}}].

\bibitem{Brivio:2017dfq}
I.~Brivio and M.~Trott, \emph{{The Neutrino Option}},
  \href{https://arxiv.org/abs/1703.10924}{{\tt 1703.10924}}.

\bibitem{Minkowski:1977sc}
P.~Minkowski, \emph{{$\mu \to e\gamma$ at a Rate of One Out of $10^{9}$ Muon
  Decays?}}, \href{http://dx.doi.org/10.1016/0370-2693(77)90435-X}{\emph{Phys.
  Lett.} {\bf B67} (1977) 421--428}.

\bibitem{GellMann:1980vs}
M.~Gell-Mann, P.~Ramond and R.~Slansky, \emph{{Complex Spinors and Unified
  Theories}}, {\emph{Conf. Proc.} {\bf C790927} (1979) 315--321},
  [\href{https://arxiv.org/abs/1306.4669}{{\tt 1306.4669}}].

\bibitem{Mohapatra:1979ia}
R.~N. Mohapatra and G.~Senjanovic, \emph{{Neutrino Mass and Spontaneous Parity
  Violation}}, \href{http://dx.doi.org/10.1103/PhysRevLett.44.912}{\emph{Phys.
  Rev. Lett.} {\bf 44} (1980) 912}.

\bibitem{Yanagida:1980xy}
T.~Yanagida, \emph{{Horizontal Symmetry and Masses of Neutrinos}},
  \href{http://dx.doi.org/10.1143/PTP.64.1103}{\emph{Prog. Theor. Phys.} {\bf
  64} (1980) 1103}.

\bibitem{Broncano:2002rw}
A.~Broncano, M.~B. Gavela and E.~E. Jenkins, \emph{{The Effective Lagrangian
  for the seesaw model of neutrino mass and leptogenesis}},
  \href{http://dx.doi.org/10.1016/j.physletb.2006.04.003,
  10.1016/S0370-2693(02)03130-1}{\emph{Phys. Lett.} {\bf B552} (2003)
  177--184}, [\href{https://arxiv.org/abs/hep-ph/0210271}{{\tt
  hep-ph/0210271}}].

\bibitem{Elgaard-Clausen:2017xkq}
G.~Elgaard-Clausen and M.~Trott, \emph{{On expansions in neutrino effective
  field theory}},  \href{https://arxiv.org/abs/1703.04415}{{\tt 1703.04415}}.

\bibitem{Brivio:2017vri}
I.~Brivio and M.~Trott, \emph{{The Standard Model as an Effective Field
  Theory}},  \href{https://arxiv.org/abs/1706.08945}{{\tt 1706.08945}}.

\bibitem{Casas:1999cd}
J.~A. Casas, V.~Di~Clemente, A.~Ibarra and M.~Quiros, \emph{{Massive neutrinos
  and the Higgs mass window}},
  \href{http://dx.doi.org/10.1103/PhysRevD.62.053005}{\emph{Phys. Rev.} {\bf
  D62} (2000) 053005}, [\href{https://arxiv.org/abs/hep-ph/9904295}{{\tt
  hep-ph/9904295}}].

\bibitem{Bambhaniya:2016rbb}
G.~Bambhaniya, P.~Bhupal~Dev, S.~Goswami, S.~Khan and W.~Rodejohann,
  \emph{{Naturalness, Vacuum Stability and Leptogenesis in the Minimal Seesaw
  Model}}, \href{http://dx.doi.org/10.1103/PhysRevD.95.095016}{\emph{Phys.
  Rev.} {\bf D95} (2017) 095016}, [\href{https://arxiv.org/abs/1611.03827}{{\tt
  1611.03827}}].

\bibitem{Aad:2015zhl}
{\scshape ATLAS, CMS} collaboration, G.~Aad et~al., \emph{{Combined Measurement
  of the Higgs Boson Mass in $pp$ Collisions at $\sqrt{s}=7$ and 8 TeV with the
  ATLAS and CMS Experiments}},
  \href{http://dx.doi.org/10.1103/PhysRevLett.114.191803}{\emph{Phys. Rev.
  Lett.} {\bf 114} (2015) 191803},
  [\href{https://arxiv.org/abs/1503.07589}{{\tt 1503.07589}}].

\bibitem{Olive:2016xmw}
{\scshape Particle Data Group} collaboration, C.~Patrignani et~al.,
  \emph{{Review of Particle Physics}},
  \href{http://dx.doi.org/10.1088/1674-1137/40/10/100001}{\emph{Chin. Phys.}
  {\bf C40} (2016) 100001}.

\bibitem{Ade:2015xua}
{\scshape Planck} collaboration, P.~A.~R. Ade et~al., \emph{{Planck 2015
  results. XIII. Cosmological parameters}},
  \href{http://dx.doi.org/10.1051/0004-6361/201525830}{\emph{Astron.
  Astrophys.} {\bf 594} (2016) A13},
  [\href{https://arxiv.org/abs/1502.01589}{{\tt 1502.01589}}].

\bibitem{Davoudiasl:2014pya}
H.~Davoudiasl and I.~M. Lewis, \emph{{Right-Handed Neutrinos as the Origin of
  the Electroweak Scale}},
  \href{http://dx.doi.org/10.1103/PhysRevD.90.033003}{\emph{Phys. Rev.} {\bf
  D90} (2014) 033003}, [\href{https://arxiv.org/abs/1404.6260}{{\tt
  1404.6260}}].

\end{thebibliography}
\end{document}